\documentclass[pre,superscriptaddress,showpacs,nofootinbib,
  floatfix,preprint,preprintnumbers]{revtex4-1}

\usepackage[utf8x]{inputenc}
\usepackage{amsfonts}
\usepackage{amsmath}
\usepackage{amssymb}
\usepackage{graphicx}
\usepackage{epstopdf}
\usepackage{hyperref}
\usepackage{bm}

\DeclareGraphicsExtensions{.pdf}

\hypersetup{colorlinks=true, citecolor=blue,
  pagecolor=blue,
  urlcolor=black,
  pdfcreator={pdflatex},
}

\begin{document}
\title{Anomalous diffusion and diffusion anomaly in confined Janus dumbbells}

\author{Leandro B. Krott}
\address{\footnotesize Centro Ararangu\'a, Universidade Federal 
de Santa Catarina, Rua Pedro Jo\~ao Pereira, 150, 
CEP 88905-120, Ararangu\'a, SC, Brazil. E-mail: leandro.krott@ufsc.br 
}

\author{Cristina Gavazzoni}
\affiliation{\footnotesize Instituto de Física, Univeridade Federal do Rio Grande do
  Sul, Caixa Postal 15051, CEP 91501-570, Porto Alegre, RS,
  Brazil.}

 \author{Jos\'e R. Bordin}
\affiliation{\footnotesize Campus Ca\c capava do Sul, Universidade Federal
do Pampa, Av. Pedro Anuncia\c c\~ao, 111, CEP 96570-000, 
Ca\c capava do Sul, RS, Brazil. Tel: +55 55 3281-1711; E-mail: josebordin@unipampa.edu.br}

\begin{abstract}
Self-assembly and dynamical properties of Janus nanoparticles have been studied by molecular dynamic simulations. 
The nanoparticles are modeled as dimers and they are confined between two flat parallel plates to simulate a 
thin film. One monomer from the dumbbells interacts by a standard Lennard Jones potential and the other
by a two-length scales shoulder potential, typically used for anomalous fluids. Here, we study the effects 
of remove the Brownian effects, typical from colloidal systems immersed in aqueous solution, and consider 
a molecular system, without the drag force and the random collisions from the Brownian motion. Self-assembly 
and diffusion anomaly are preserved in relation to the Brownian system. Additionally,
a superdiffusive regime associated to a collective reorientation in a highly structured phase
is observed. Diffusion anomaly and anomalous diffusion are explained in the two length 
scale framework.

\end{abstract}
\maketitle

\section{Introduction}
\label{Intro}

Janus nanoparticles are characterized as particles with two or more surfaces with distinct chemical
or physical properties. They have distinct shapes, as spheres, rods, disks and dumbbells, and
can assemble in a variety of nanoscale morphologies like spheres, cylinders, 
lamellae phases and micelles with distinct shapes and functionality~\cite{Roh05, Klapp16, Li12, White10, Munao13, Munao14, 
Munao15b, Avvisati14, Bordin16a, BoK15c}. Due it versatility, Janus nanoparticles have
promising application to medicine, self-driven molecules, catalysis, photonic crystals, stable emulsion and self-healing 
materials~\cite{Cas89, Talapin10,ElL11, TuP13, WaM08, WaM13, Zhang15, Bic14, Ao15}.
Particularly, Janus dimers~\cite{Yin01, SiC14, Lu02, YoL12} are nanoparticles formed by two monomers
linked together. Each monomer has distinct functionality and characteristic, as well can have
different size.

Regarding to diffusion, we can rise two interesting phenomena. The first is the anomalous diffusion.
The diffusion coefficient is obtained from the scaling factor between the mean square displacement,  $<r(0)r(t)>$,
and the exponent of time, $t^{\alpha}$. For regular, or Fick, diffusive process, $\alpha = 1.0$.
If $\alpha > 1.0$ we say that the system is superdiffusive, and if $\alpha < 1.0$ the regime is subdiffusive.
For Brownian systems, the collisions between solute and solvent and the solvent viscosity,
considered using a white noise and a drag force~\cite{AllenTild}, usually lead the system
to a Fick diffusion. On the other hand, for molecular systems, where there is no
solvent effects, the Fick diffusion is achieved due the collisions between 
the particles of the system. If the collision are rare, as in a infinitively diluted gas,
the diffusion tends to be ballistic. Anomalous diffusion was observed in several systems, as 
proteins and colloids in crowded environments~\cite{Banks05, Illien14},
self-propelled particles~\cite{Hagen11} and confined nanoparticles~\cite{Xue16}.

The second interesting phenomena is the diffusion anomaly. For most materials,
the diffusion coefficient decreases when the pressure (or density)
increases. However, materials as water~\cite{Ne02a}, silicon~\cite{Mo05} and silica~\cite{Sa03} 
show diffusion anomaly, characterized by a
maximum in the diffusion coefficient at constant temperature. Besides diffusion (or dynamical) anomaly,
water, silicon, silica and others fluids, the so-called anomalous fluids, also have
other classes of anomalies, as structure and thermodynamic anomalies.

Experimental studies have reported the production of 
silver-silicon (Ag-Si)\cite{SiC14} and silica-polystyrene (SiO$_2$-PS)\cite{Liu09}
dimeric nanoparticles. Therefore, we have a Janus dumbbell with one anomalous monomer and another regular, or non-anomalous,
monomer. Inspired in this specific shape and composition, we have proposed a model to study this class of
nanoparticles~\cite{BoK15c, Bordin16a} based in a effective two length scales potential~\cite{Ol06a} 
to model the anomalous monomer, while the regular monomer is modeled with a standard Lennard-Jones (LJ)
potential. We have shown that despite the presence of the non-anomalous monomer, the diffusion anomaly was preserved.

Usually, colloidal solutions are dissolved in a solvent - as water. In these cases, Langevin Dynamics has 
been employed to mimic the solvent effects in the colloids~\cite{Bordin16a, BoK15c, BoK16a}.
On the other hand, molecular systems, without solvent effects, are relevant as well. For instance, 
Muna\`o and Urbic recently proposed a model for alcohols combining anomalous and regular monomers~\cite{Munao15c}
and Ubirc proposed a model for methanol~\cite{Hus15}. 
As well, anomalies in dimeric anomalous systems has been object of computational studies~\cite{Munao16, Ga14, Ol10}.

Therefore, a natural question that rises is how the Janus dimer will behave when the Brownian dynamics effects are removed.
In this way, we perform intensive Molecular Dynamics (MD) simulations using the Berendsen thermostat.
Comparisons are made with the system behavior in our previous work~\cite{BoK16a}, where
we have used the Langevin thermostat. The dimers are confined between two parallel
walls, in order to simulate a thin film. We show how the thermostat affects the aggregation,
the dynamic and the thermodynamic properties of this Janus nanoparticle.
Specially, we show how the absence of solvent effects 
leads the system to have not only diffusion anomaly, but also
a superdiffusive regime related to a highly ordered structure.

The paper is organized as follows: first we introduce the model and describe the methods 
and simulation details; next the results and discussion are given; and 
then we present our conclusions.

\section{The Model and the Simulation details}
\label{Model}

\begin{figure}[ht]
\begin{center}
\includegraphics[width=12cm]{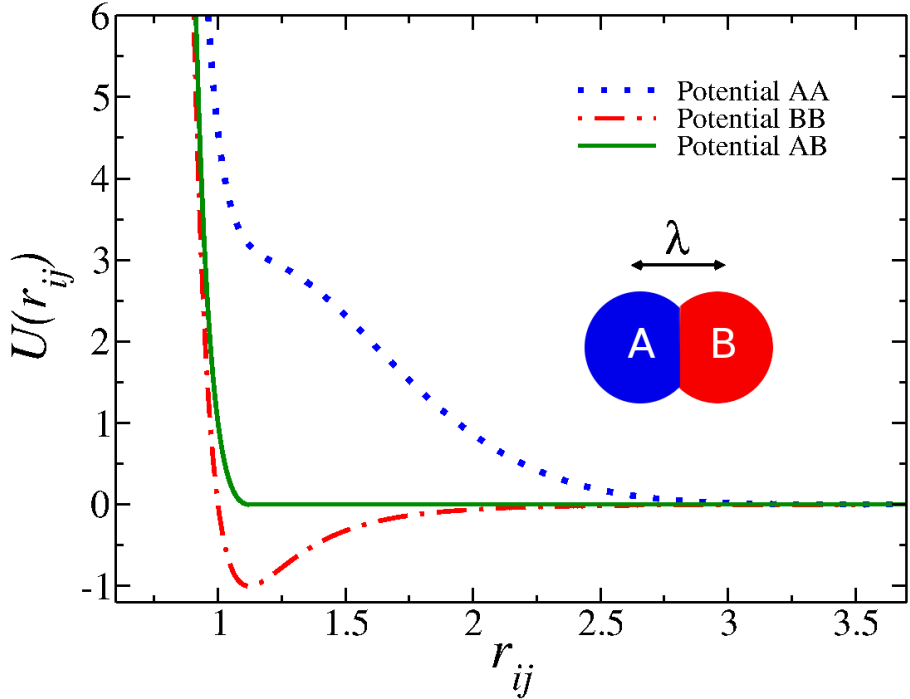}
\end{center}
\caption{Interaction potentials between particles of type A (potential AA in dotted 
blue line), between particles of type B (potential BB in dot-dashed red line) and between particles of type A 
and B (potential AB in solid green line). Potential AB in z-direction gives the interaction between dimers and walls. The inset shows 
a picture of a Janus dumbbells formed by particles A and B, with monomers separated by a distance $\lambda$.}
\label{fig1}
\end{figure}

The system consists in Janus dumbbells confined between two flat and parallel plates 
separated by a fixed distance in z-direction. The Janus particles  
are formed by $N = 288$ dimers, totalizing $N = 576$ monomers, linked 
rigidly at a distance $\lambda = 0.8\sigma$. The monomers can be of type A and type B.
Particles of type A interact through a two length scales shoulder potential, defined as~\cite{Ol06a, Ol06b}

 $$
 \frac{U^{AA}(r_{ij})}{\varepsilon} = 4\left[ \left(\frac{\sigma}{r_{ij}}\right)^{12} -
 \left(\frac{\sigma}{r_{ij}}\right)^6 \right] + 
 $$
 \begin{equation}
 u_0 {\rm{exp}}\left[-\frac{1}{c_0^2}\left(\frac{r_{ij}-r_0}{\sigma}\right)^2\right]
 \label{AlanEq}
 \end{equation}

\noindent where $r_{ij} = |\vec r_i - \vec r_j|$ is the distance between two A particles $i$ and $j$. The first term
is a standard 12-6 Lennard-Jones (LJ) potential~\cite{AllenTild} and the second one is a Gaussian shoulder centered at $r_0$, 
with depth $u_0$ and width $c_0$. The parameters used are $u_0 = 5.0$, $c_0 = 1.0$ and $r_0/\sigma = 0.7$. 
In figure~\ref{fig1}, this potential is represented by potential AA. Monomeric and dimeric bulk systems modeled by this 
potential present thermodynamic, dynamic and structural anomalies
like observed in water, silica and other anomalous fluids~\cite{Ol06a, Ol06b, Ol10, Kell67,Angell76}. 
Under confinement, the monomeric system also presents water-like anomalies and interesting 
structural phase transitions~\cite{Krott13,Krott14,BoK14c,Krott13b,Bordin14a,BoK15a,Bor12,Bordin13a,Bordin14b, KoB15}.

Particles of type B interact through a standard 12-6 LJ potential, whose equation is the same of first 
term of Eq.~\ref{AlanEq}. This potential is represented in figure~\ref{fig1} by potential BB. We use a 
cutoff radius of $r_c = 2.5$. Finally, the interaction between particles of type A and type B is given 
by a Weeks-Chandler-Andersen (WCA) potential, defined as~\cite{AllenTild}
 
 \begin{equation}
 \label{LJCS}
 U^{\rm{CSLJ}}(r_{ij}) = \left\{ \begin{array}{ll}
 U_{{\rm {LJ}}}(r_{ij}) - U_{{\rm{LJ}}}(r_c)\;, \qquad r_{ij} \le r_c\;, \\
 0\;, \qquad \qquad \qquad \qquad \quad r_{ij}  > r_c\;.
 \end{array} \right.
 \end{equation}

\noindent The WCA potential considers just the repulsive part of the standard 12-6 LJ potential, as shown in 
figure~\ref{fig1} by potential AB. Dimers and walls also interact by a WCA potential, but projected in the $z$-direction,

\begin{equation}
 \label{wall}
 U^{\rm{wall}}(z_{ij}) = \left\{ \begin{array}{ll}
 U_{{\rm {LJ}}}(z_{ij}) - U_{{\rm{LJ}}}(z_c)\;, \qquad z_{ij} \le z_c\;, \\
 0\;, \qquad \qquad \qquad \qquad \quad z_{ij}  > z_c\;.
 \end{array} \right.
 \end{equation}
\noindent here, $z_c = r_c = 2^{1/6}\sigma$.

The walls are fixed in $z$-direction separated by a distance of $L_z = 4\sigma$. We choose this separation 
to observe the effects of strong confinement, like in quasi-2D thin films. Standard periodic boundary 
conditions are applied in $x$ and $y$ directions. 

The simulations are done in NVT-ensemble using a home made program. The temperature was fixed with the
Berendsen thermostat and the equation motions were integrated by velocity Verlet algorithm with a 
time step $\delta t = 0.01$ in reduced unit. The number density is calculated as $\rho = N/V$, where $N$ is 
the number of monomers particles and $V=L^2L_z$ is the volume of the simulation box. We simulate systems 
varying the values of $L$ from $19.0\sigma$ to $60.0\sigma$ to obtain different densities. We use 
the SHAKE algorithm~\cite{Ryc77} to link rigidly each dimer at a distance of $\lambda = 0.8\sigma$.

We performed $5\times10^5$ steps to equilibrate the system followed by $5\times10^6$ steps run 
for the results production stage. The equilibrium state was checked by the analysis of potential 
and kinetic energies as well the snapshots of distinct simulation times. Simulations with up to 
$N = 5000$ particles were carried out, and essentially the same results were obtained.

The dynamic of the system was analyzed by the lateral mean square displacement and the  velocity autocorrelation 
function $v_{acf}$. The lateral mean square displacement was calculated by

\begin{equation}
\label{r2}
\langle [\vec r_{\parallel\rm cm}(t) - \vec r_{\parallel\rm cm}(t_0)]^2 \rangle =\langle \Delta \vec r_{\parallel\rm cm}(t)^2 \rangle\;,
\end{equation}

\noindent where $\vec r_{\parallel\rm cm}(t_0) = (x_{\rm cm}(t_0)^2 + y_{\rm cm}(t_0)^2)$ 
and  $\vec r_{\parallel\rm cm}(t) = (x_{\rm cm}(t)^2 + y_{\rm cm}(t)^2)$
denote the parallel coordinate of the nanoparticle center of mass (cm)
at a time $t_0$ and at a later time $t$, respectively. Fick diffusion has a diffusion coefficient that
can be calculated by Einstein relation,

\begin{equation}
 D_{\parallel} = \lim_{t \rightarrow \infty} \frac{\langle \Delta \vec r_{\parallel\rm cm}(t)^2 \rangle}{4t}\;.
\end{equation}

The velocity autocorrelation function $v_{acf}$ was evaluated taking the average for all nanoparticles 
center of mass and initial times
 \begin{equation}
 \label{vacf}
 v_{acf} = <\vec v_i(t_0)\cdot \vec v_i(t+t_0)>\;,
 \end{equation}

\noindent where $\vec v_i(t_0)$ is the initial velocity for the $i$-th nanoparticle center of mass and 
$\vec v_i(t+t_0)$ is the velocity at an advanced time $t+t_0$ for the same particle center of mass.

The system structure was analyzed with the lateral radial distribution function $g_{||}(r_{||})$, evaluated
in the $xy$ plane in all phases and defined as~\cite{Ku05b}

\begin{equation}
\label{gr_lateral}
g_{||}(r) \equiv \frac{1}{\rho ^2V}
\sum_{i\neq j} \delta (r-r_{ij}) \left [ \theta\left( \left|z_i-z_j\right| + \frac{\delta z}{2}
\right) - \theta\left(\left|z_i-z_j\right|-\frac{\delta z}{2}\right) \right],
\end{equation}
\noindent where $\delta(x)$ is the Dirac $\delta$ function
and the Heaviside function $\theta (x)$ restricts the sum of particle pair in the same
slab of thickness $\delta z = \sigma$. 

Directly related to $g_{||}(r_{||})$, we also use the 
translational order parameter $\tau$, defined as~\cite{Er01}
\begin{equation}
\label{order_parameter}
\tau \equiv \int^{\xi _c}_0  \mid g_{\parallel}(\xi)-1  \mid d\xi,
\end{equation}
\noindent where $\xi = r_{\parallel}(\rho^{l})^{1/2}$ is the interparticle 
distance in the direction parallel to the plates scaled by the density of the layer,
$\rho^{l} = N^{l}/L^2$. $N^{l}$ is the average of particles for each layer. 
We use $\xi_c = (\rho^{l})^{1/2}L/2$ as cutoff distance. 

The snapshots of the systems also were used to analyze the self-assembled structures at
different densities and temperatures. The pressure-temperature phase diagram was 
constructed using the parallel pressure ($P_{\parallel}$), calculated by Virial expression in $x$ and $y$ directions.
All the system properties were evaluated for all simulated points.

All physical quantities are computed in the standard LJ units\cite{AllenTild}. Distance, density of particles, time,
parallel pressure and temperature are given, respectively, by

\begin{equation}
\label{red1}
r^*\equiv \frac{r}{\sigma}\;,\quad \rho^{*}\equiv \rho \sigma^{3}\;, \quad 
\mbox{and}\quad t^* \equiv t\left(\frac{\epsilon}{m\sigma^2}\right)^{1/2}\;, \quad
P_{\parallel}^*\equiv \frac{p \sigma^{3}}{\epsilon} \quad \mbox{and}\quad 
T^{*}\equiv \frac{k_{B}T}{\epsilon}\; ,
\end{equation}

\noindent where $\sigma$, $\epsilon$ and $m$ 
are the distance, energy and mass parameters, respectively.
Considering that, we will omit the symbol $^*$ to simplify the discussion.

\section{Results and Discussion}
\label{Results}

\begin{figure}[!htp]
 \begin{center}
  
 \includegraphics[width=5cm]{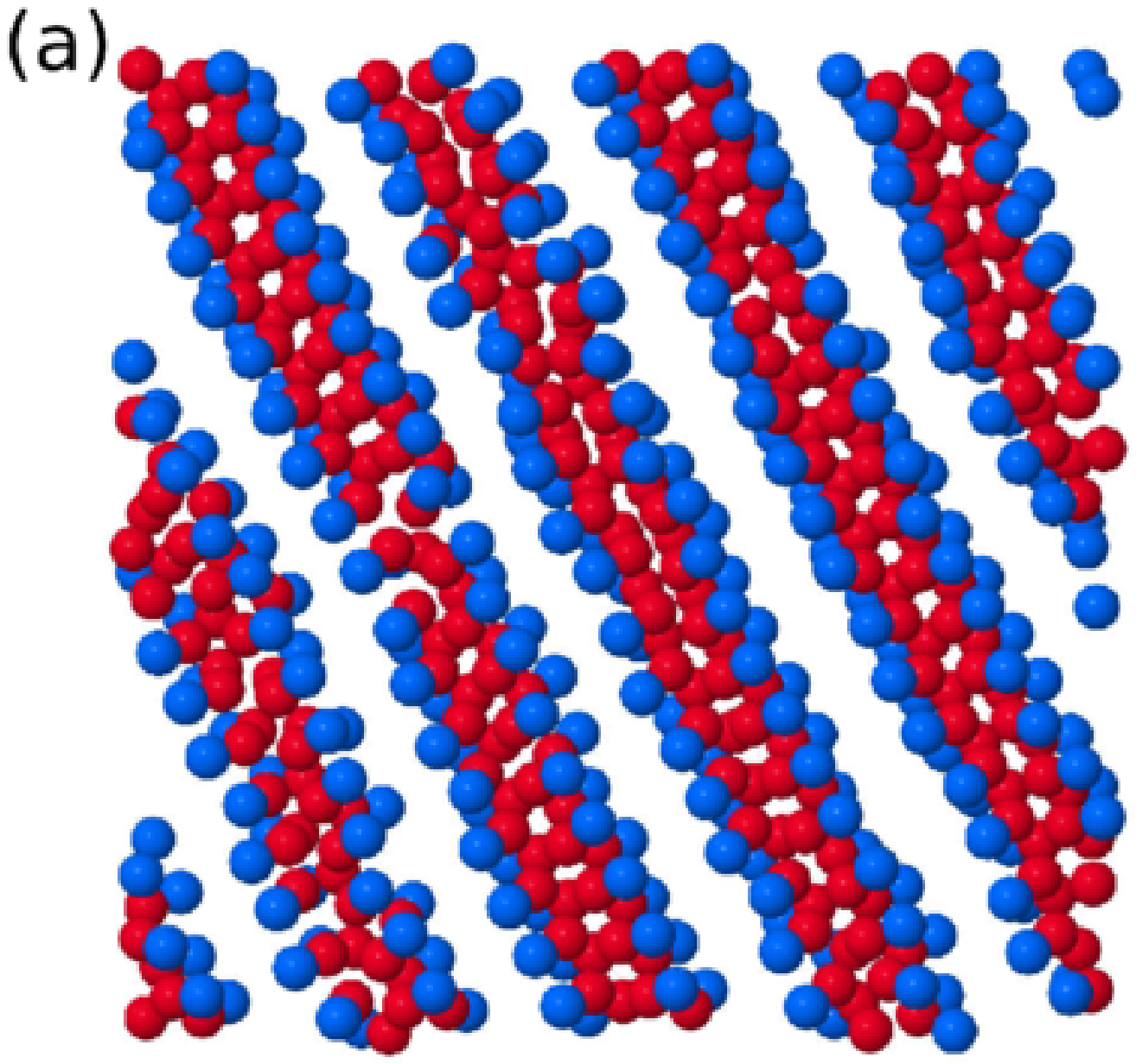}
 \includegraphics[width=5cm]{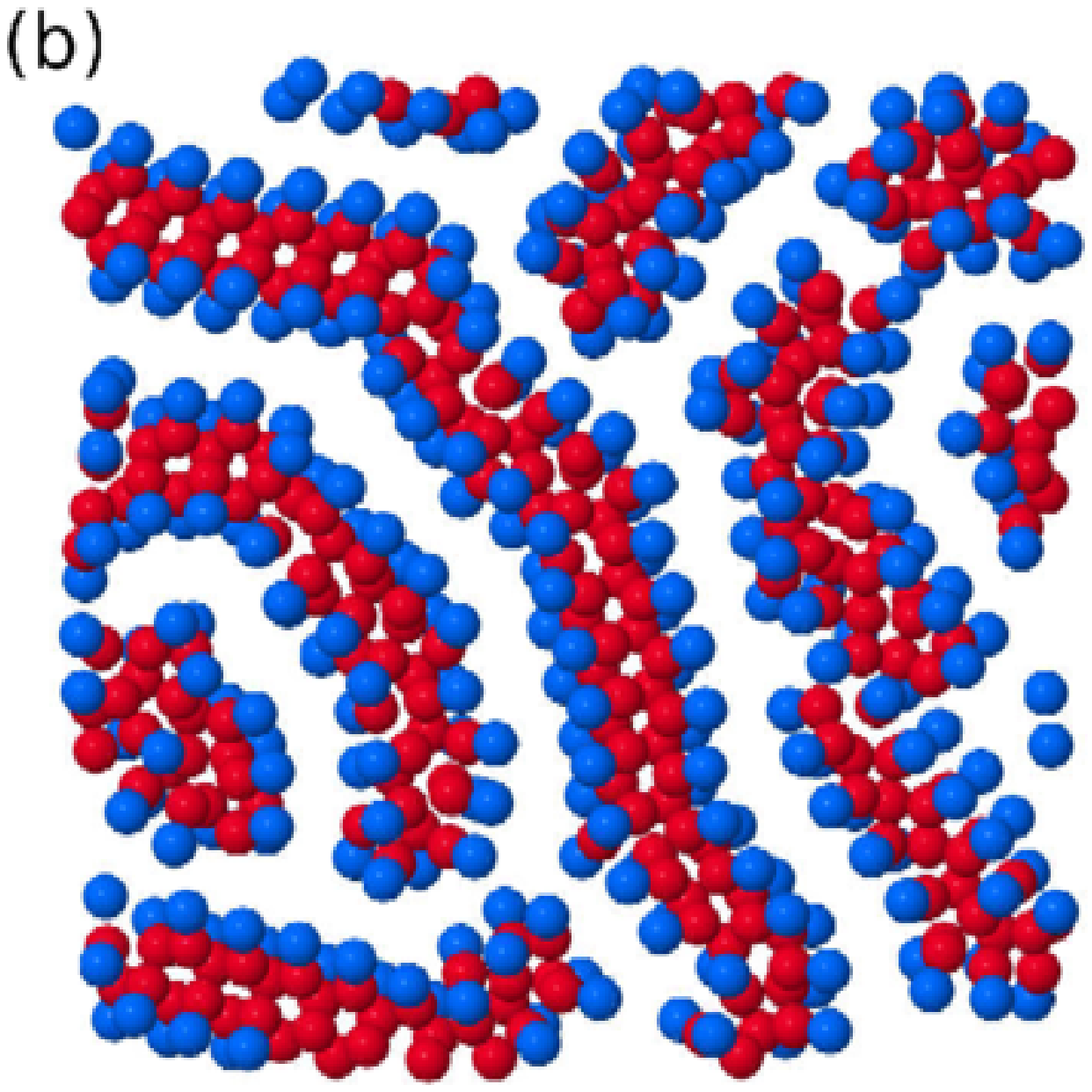}
 \includegraphics[width=5cm]{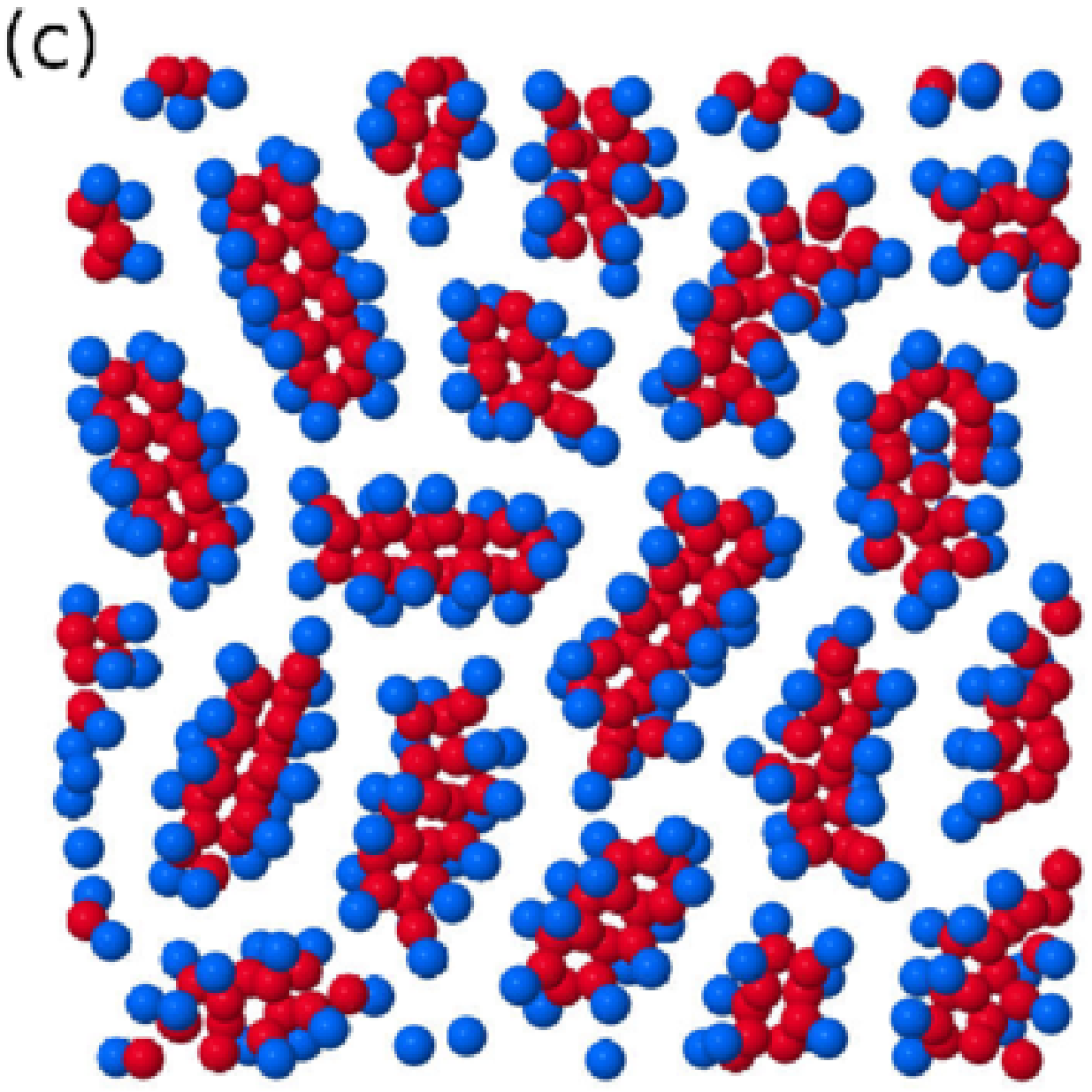}
 \includegraphics[width=5cm]{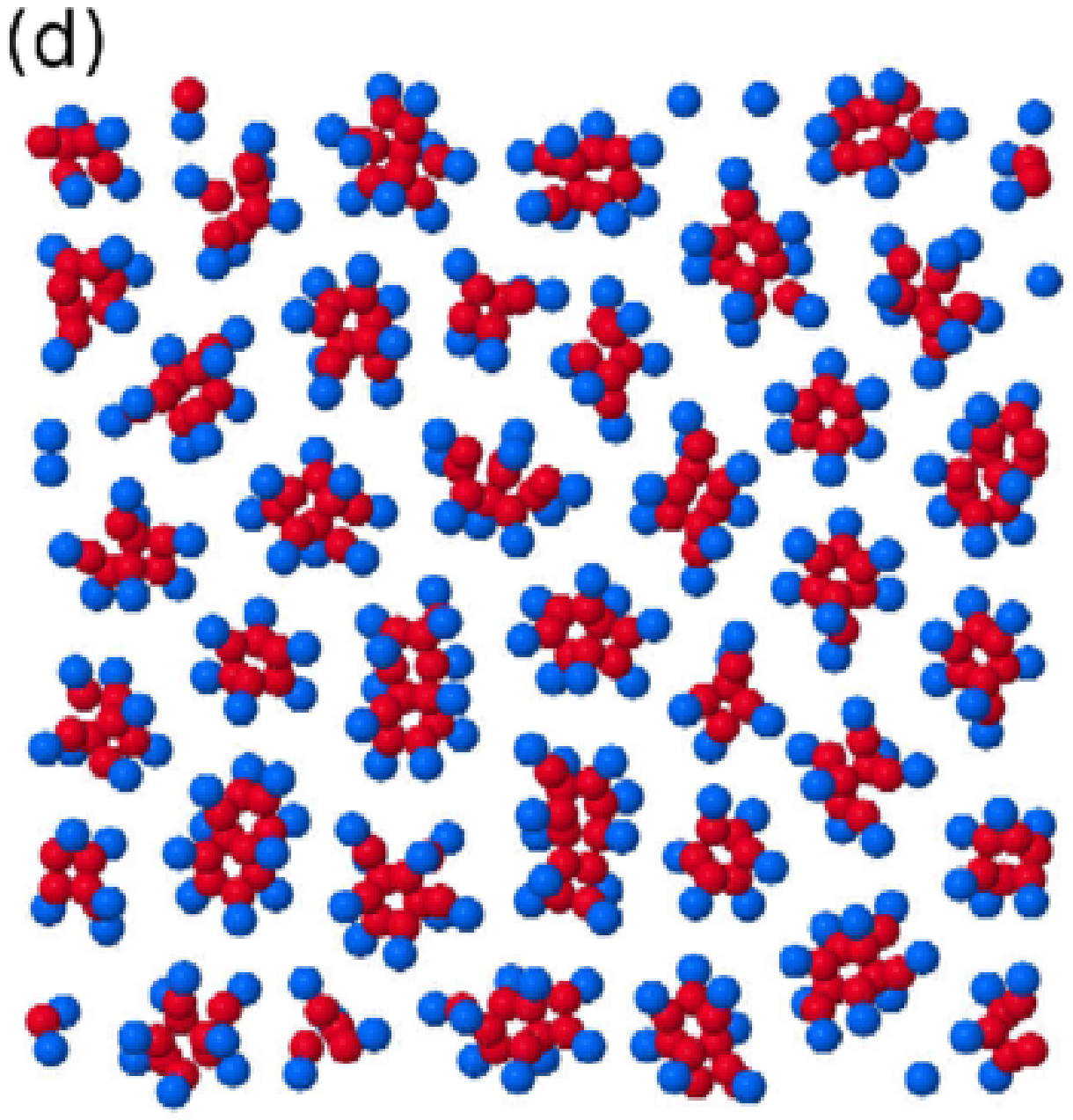}
 \includegraphics[width=5cm]{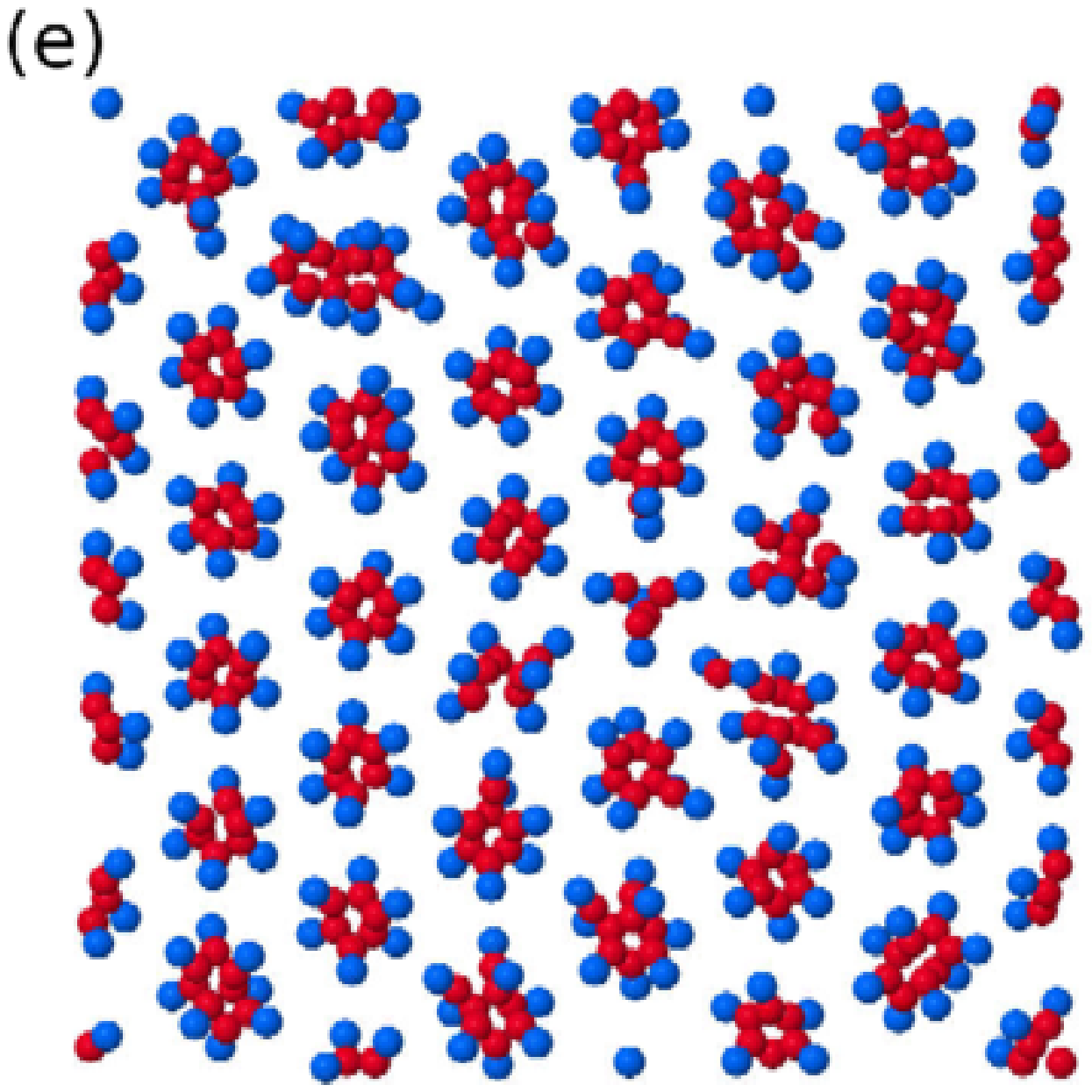}
 \includegraphics[width=5cm]{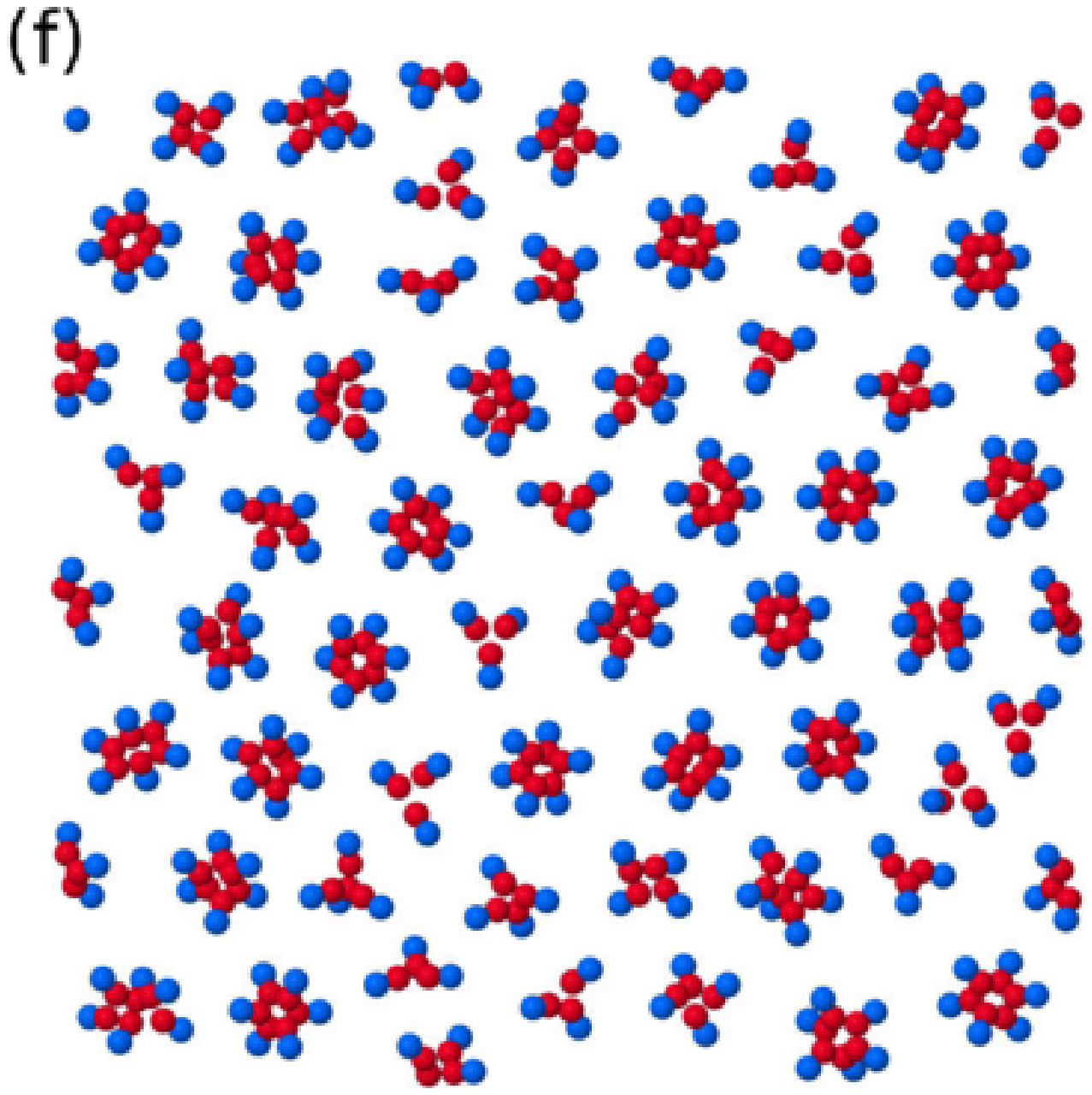}
 \includegraphics[width=5cm]{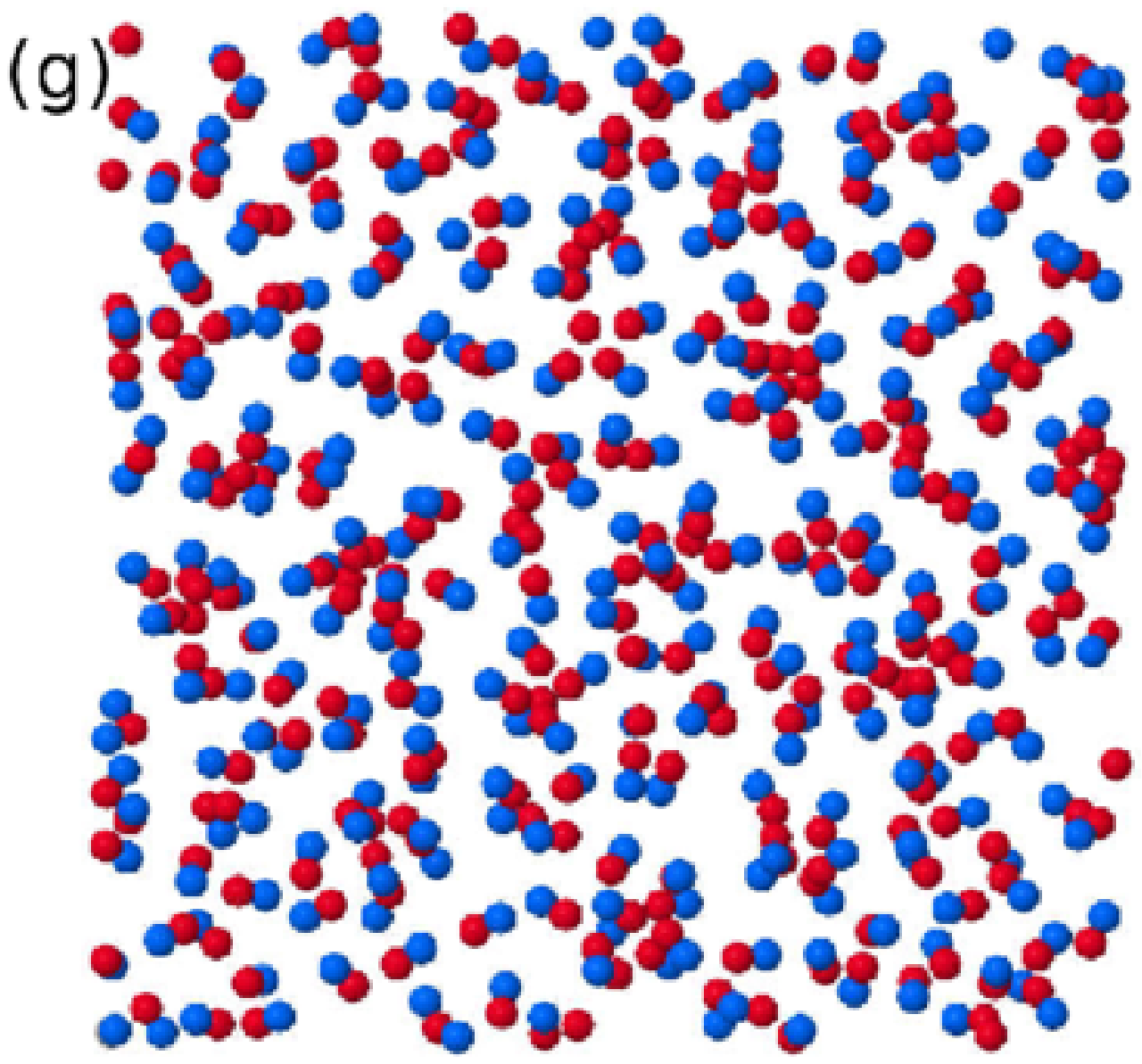}
 \includegraphics[width=5cm]{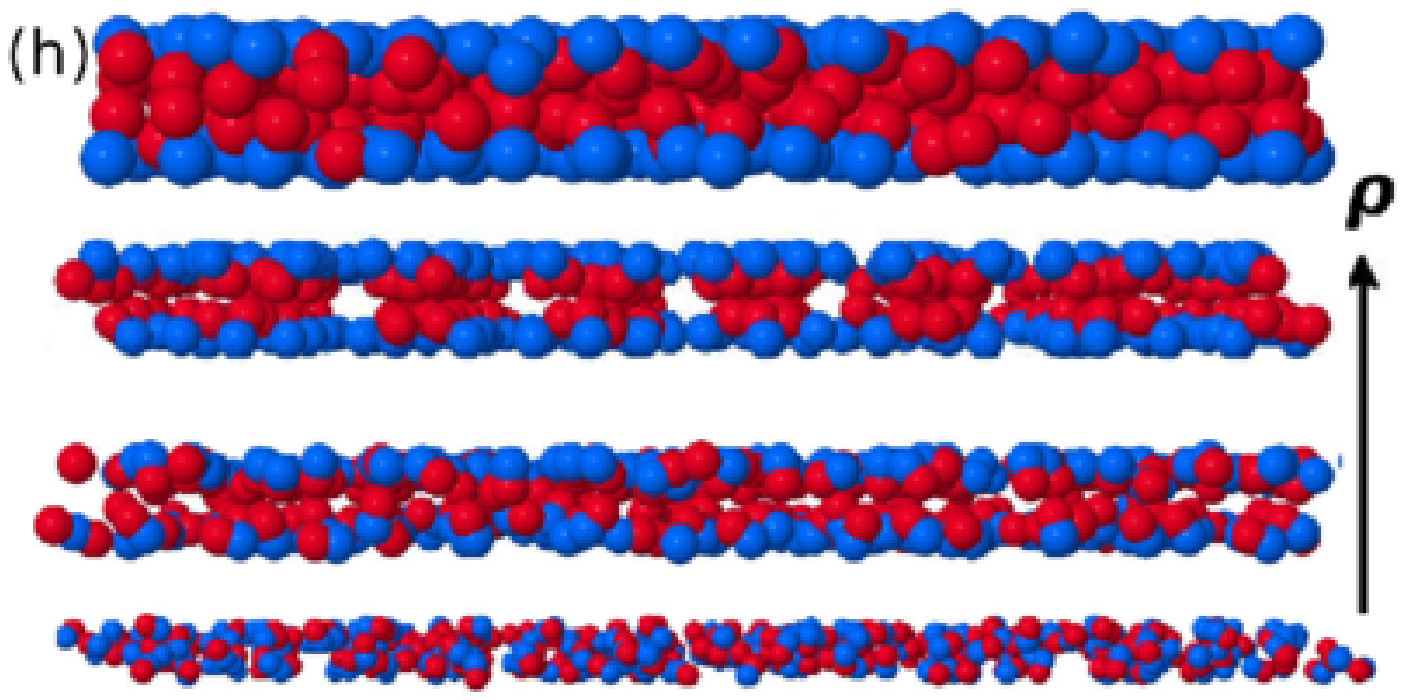}
 \end{center}
 \caption{Frontal vision snapshots of confined Janus particles at $\lambda = 0.8$. We 
 show (a) Structured elongated micelles (SEM) at $\rho = 0.48$ and $T = 0.200$, (b) 
elongated micelles I (EMI) at $\rho = 0.50$
 and $T = 0.100$, (c)elongated micelles II (EMII) at $\rho = 0.40$ and $T = 0.100$, (d) 
 spherical micelles (SM) at $\rho = 0.28$ and and $T = 0.100$, (e) hexagonal aggregates (HA) at 
$\rho = 0.25$ and and $T = 0.100$, 
 (f) coexistence between hexagonal and trihedral aggregates (TA) at $\rho = 0.15$ and and $T = 0.100$,
 (g) fluid phase at $\rho = 0.25$ and $T = 0.400$. (h) Lateral view of the simulation box showing the 
 bilayer structure for smaller (bottom) and higher (up) densities. 
 For simplicity, the plates are not shown.}
 \label{snaps}
 \end{figure}

Controlling the self-assembly of chemical building blocks
in distinct structures is the goal in the studies of confined
molecules and nanoparticles with this characteristic~\cite{Kim15, Ro12, Ro11}, as Janus nanoparticles.
Therefore, we first report the self-assembled structures observed in our 
simulations. The aggregates were classified using snapshots, the lateral mean square displacement, the $v_{acf}$ and the $g_{||}(r_{||})$.

The small separation $L_z = 4.0$ between the plates used in our simulations induces the system to
form two layers, one near each plate, regardless the system density, as we shown in figure~\ref{snaps}(h). 
To map the particles arrangement, we analyze the frontal vision of the system, summarized in figure~\ref{snaps}.
At high temperatures, the system is in a fluid phase, without a defined arrangement.
For instance, we can see the snapshot at density $\rho = 0.25$ and temperature $T = 0.400$,
shown in figure~\ref{snaps}(g). At lower temperatures, the nanoparticles aggregate.
The shape and structure of these aggregates depend on the system density. 
Let's take the isotherm $T = 0.100$. At the smallest densities, $\rho < 0.16$,
the system remains in a fluid phase. The aggregation starts at $\rho = 0.16$, with
the nanoparticles assembled in trihedral (TA3), tetrahedrical (TA4) or hexagonal (HA) aggregates.
As the name indicates, the first one is composed of three dimers, the second one
by four and the third by six dimers. The snapshot in figure~\ref{snaps}(f) shows the coexistence of
these aggregates. Increasing the density, the TA3/4 binds and more HA aggregates are observed,
as shown in figure~\ref{snaps}(e) for $\rho = 0.25$.

Spherical micelles (SM) were also obtained, as we show in figure~\ref{snaps}(d)
for $\rho = 0.28$, as well elongated micelles (EMII), shown in figure~\ref{snaps}(c).
HA, SM and EMII structures were observed in the bulk system~\cite{BoK15c}. The confinement
leaded to the assembly in TA3/4 at lower densities 
and longer elongated micelles (EMI) at higher densities. The EMI can
show no preferential orientation, as the one shown in figure~\ref{snaps}(b) for $\rho = 0.50$
and $T = 0.100$, or have a well defined orientation, as in the figure~\ref{snaps}(a)
for $\rho = 0.48$ and $T = 0.200$. We will refer to this last case of oriented elongated micelles as
structured elongated micelles (SEM). The main difference in the aggregation compared to the
Brownian system~\cite{BoK16a} is in the SEM phase. Here, this lamellar phase is 
straight, while in our previous work we have observed a rippled lammelae structure.
Nevertheless, this small structural difference leads to a new dynamical feature,
as we discuss below.

\begin{figure}[!htp]
 \begin{center}
 \includegraphics[width=8cm]{fig3a.eps}
 \includegraphics[width=8cm]{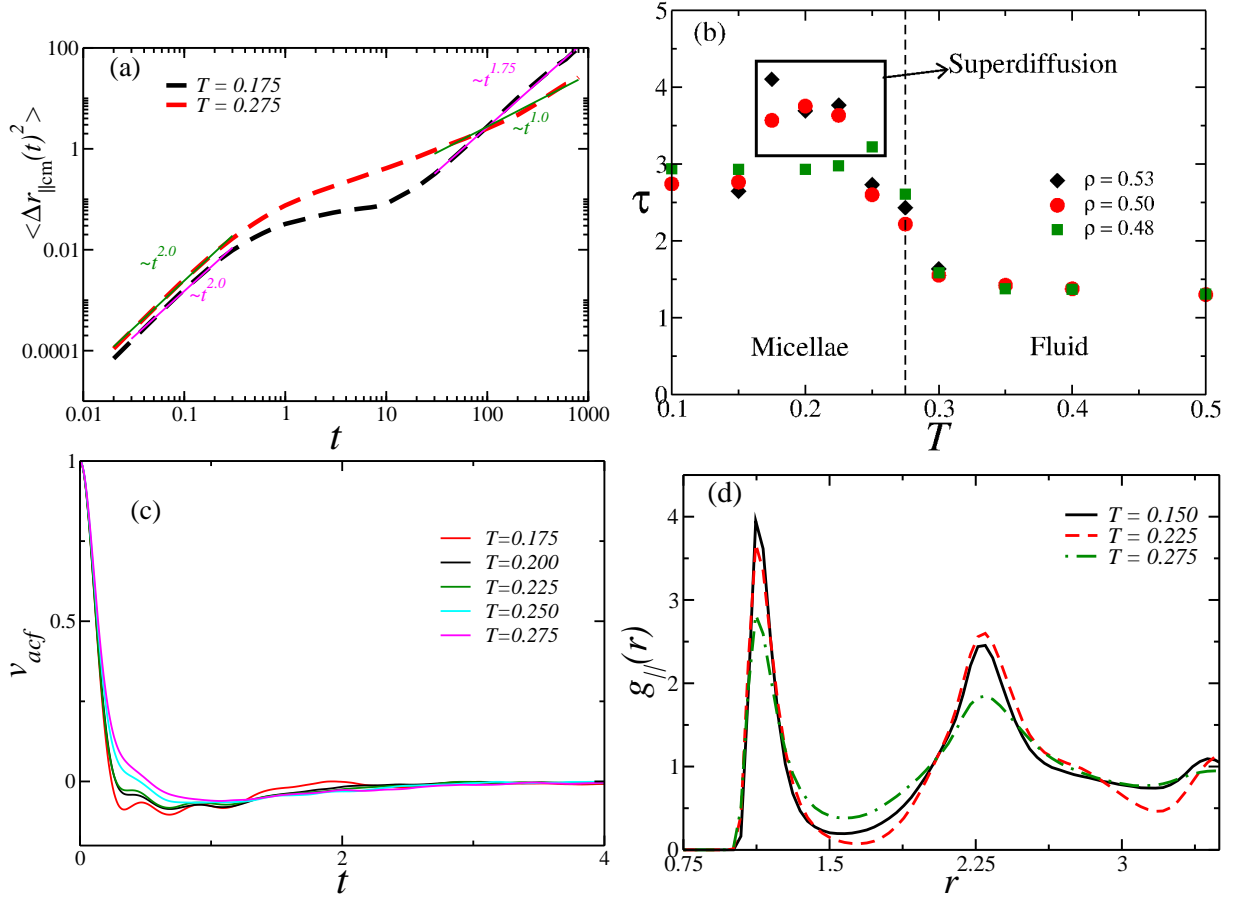}
  \includegraphics[width=8cm]{fig3c.eps}
  \includegraphics[width=8cm]{fig3d.eps}
 \end{center}
 \caption{(a) Lateral mean square displacement as function of time for nanoparticles center of mass in the
 Fick (red dashed line) and superdiffusive (black dashed line) regime. Straight lines in green 
 and magenta are guide to the eyes to show the curves slope. (b) Translational order parameter
 $\tau$ as function of temperature for distinct isochores showing the relation between the maximum
 in $\tau$ and the superdiffusivity. (c) Velocity autocorrelation function for $\rho = 0.53$ and distinct velocities.
 (d) Lateral radial distribution function showing the behavior of the first two peaks, related to the
 two length scales.}
 \label{compare}
 \end{figure}

The structural and dynamical behavior of two length scale fluids are strongly connected~\cite{Ol06a, Ol06b, Bordin13a}.
Therefore, in order to understand the difference between the EMI and SEM aggregates
we investigated the dynamical characteristics of these micelles. For instance, we take the isochore
$\rho = 0.53 $, where the Janus dimers are assembled in both EMI and SEM . The lateral mean square displacement for temperature $T=0.275$, 
red dashed line in figure~\ref{compare}(a), shows an initial ballistic behavior, and a 
Fick regime as $t\rightarrow \infty$.
On the other hand, for $T=0.175$, black dashed curve in figure~\ref{compare}(a), the initial ballistic
evolves to a superdiffusive regime, with $\alpha \approx 1.75$. This superdiffusive regime was observed
in all SEM structures, with $\alpha$ ranging from $\sim1.30$ to $\sim1.80$. 
For simplicity, we will not show all lateral mean square displacement curves.
Basically, in the fluid regions and for all aggregates, except the SEM, the nanoparticles collisions lead
the system to the Fick regime. However, in the SEM lamellae phase, the liquid-crystal structure 
prevents the collisions. As consequence, the dimers moves freely in the line defined by the structure,
leading to the superdiffusive regime. For the Brownian system~\cite{BoK16a}, since 
the lamellae phase is not straight, but rippled, the supperdifusive regime was not 
observed. This shows how that the white noise and drag force from the Brownian Dynamics
effects on the dynamic will reflect in the structure.
Particularly, the SEM region corresponds to a region where the translational order
parameter $\tau$ has a maximum. This result, shown in figure~\ref{compare}(b), highlights the relation
between structure and dynamics. Therefore, the SEM micelles are highly diffused and highly structured aggregates.
This relation between structure and dynamics is well known in the literature of anomalous fluids~\cite{Ol06a, Ol06b, Bordin13a}.
However, was never relate to anomalous diffusion, only to diffusion anomaly.

The velocity autocorrelation function $v_{acf}$ is a powerful tool to understand the dynamics of the systems.
In figure~\ref{compare}(c) we show $v_{acf}$ as function of time for $\rho = 0.53$ and temperatures ranging from 
$T=0.175$ to $T=0.275$. The superdiffusive regime and SEM micelles were observed for $T=0.175$, 
$0.200$ and $0.225$. As we can see in
figure~\ref{compare}(c), at these temperatures the curves cross the zero axis at shorter times than
in the Fick regime with EMI micelles, $T = 0.250$ and $0.275$. This shows that in the superdiffusive regime the nanoparticles are
strongly caged by the neighbors dimers. Thereat, we know that the system has a superdiffusion regime
related to a collective behavior and assembly in a specific well structured micelle. 

It is well reported in the literature that the competition between 
the two scales in the Eq.~(\ref{AlanEq}) is the main ingredient
to a fluid present water-like anomalies~\cite{Oliveira07}. To see the influence of this competition, we 
analyzed the $g_{||}(r_{||})$ for the A monomers when the system enters and leaves the SEM phase. 
As we show in figure~\ref{compare}(d), when we walk
through the isochore  $\rho = 0.53$, from $T=0.150$ -- before the superdiffusion region -- to $T=0.225$
-- the limit of the superdiffusion region -- the first peak in the $g_{||}(r_{||})$ decays, while the second peak
rises. This is the competition between the scales, where the particles moves from one of the preferential
distances to the other. As we heat the system, both peaks decays, as when we walk from $T=0.225$ to
$T=0.275$ -- outside the superdiffusive regime. Therefore, the competition between the two length scales
leads the nanoparticles from a non-oriented Fick diffusion elongated micellae phase to a 
oriented superdiffusive elongated micellae phase.

  \begin{figure}[!htp]
 \begin{center}
 \includegraphics[width=8cm]{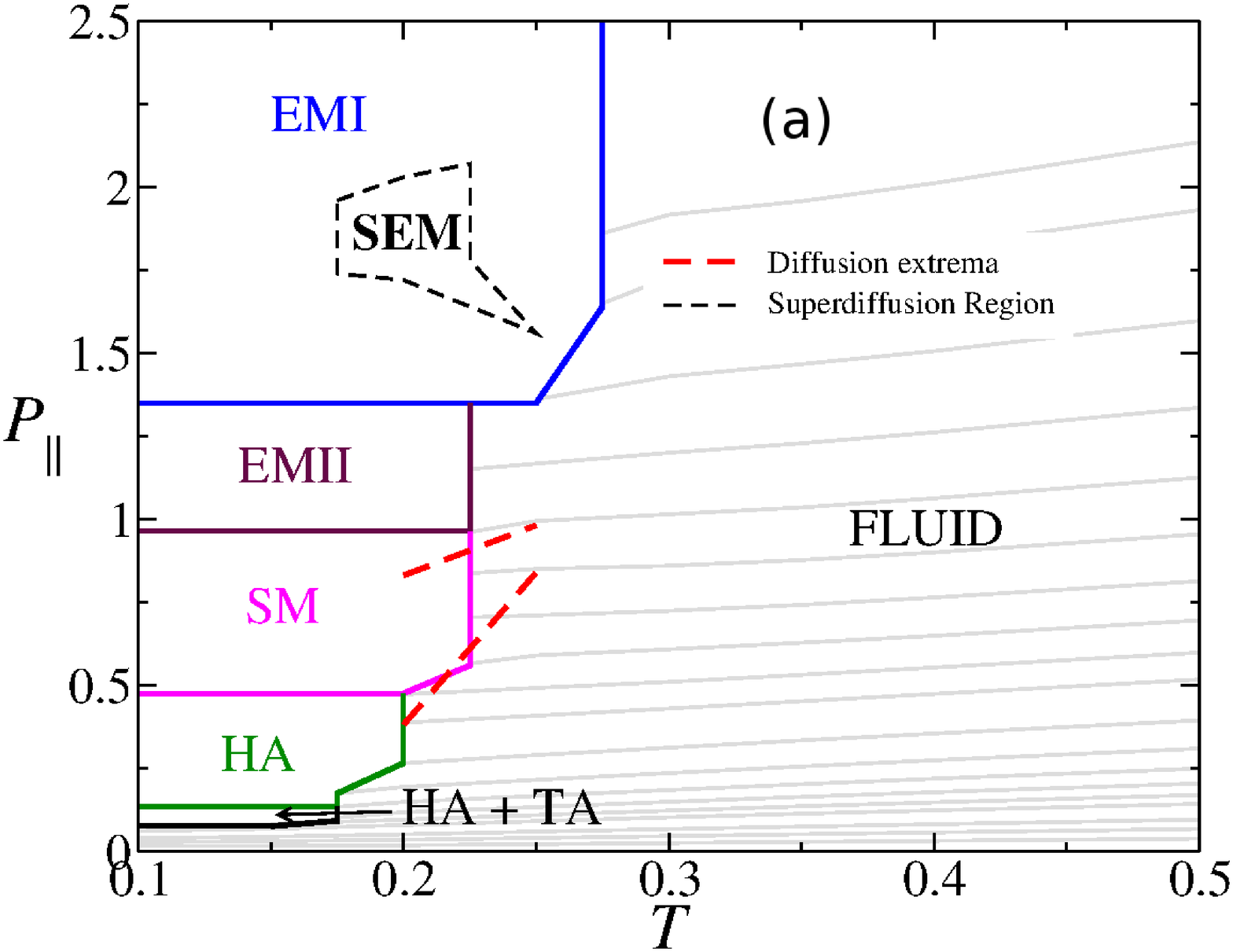}
 \includegraphics[width=8cm]{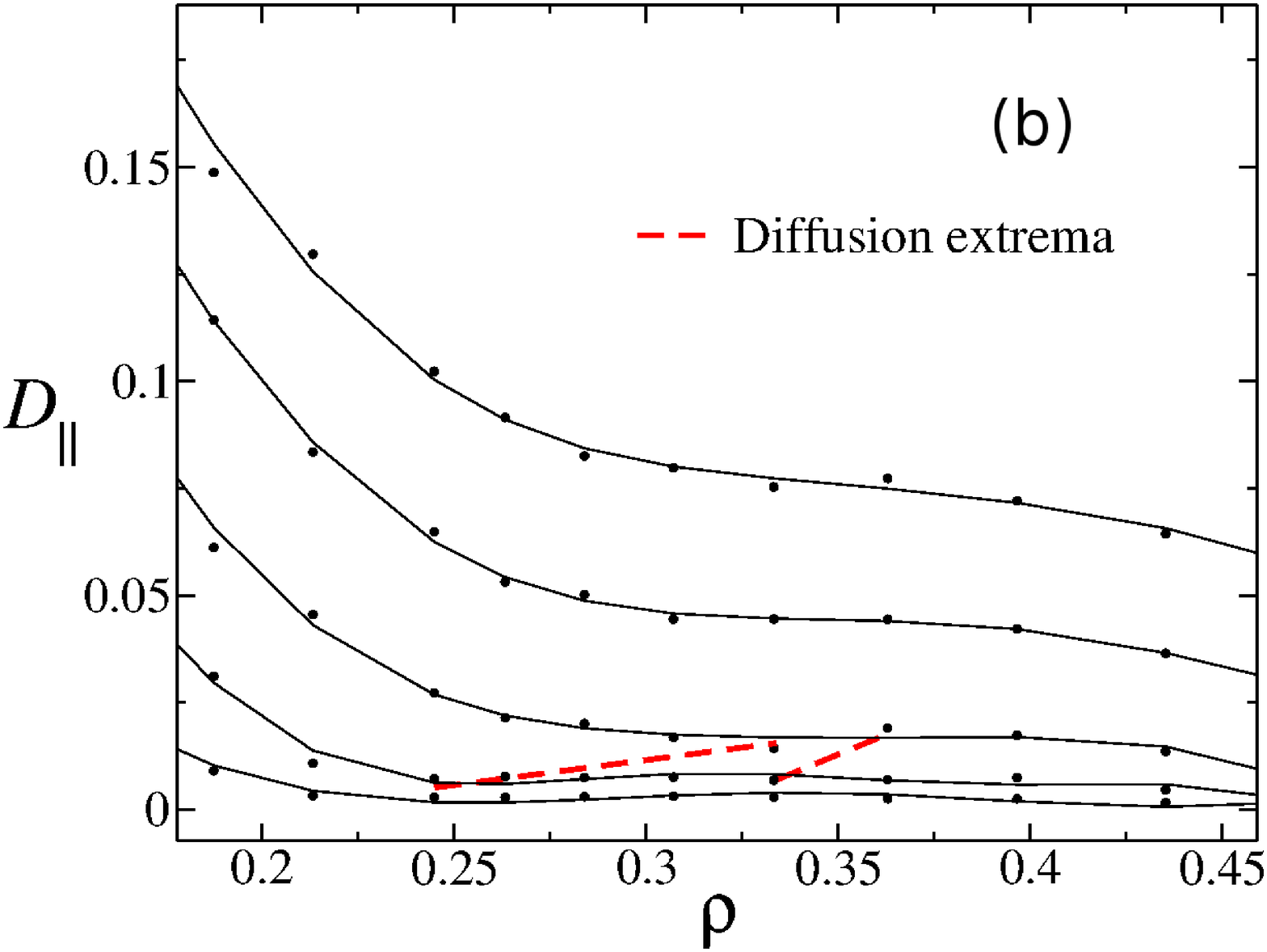}
 \end{center}
 \caption{Results for Janus confined particles with $\lambda = 0.8$. In (a) we show the parallel pressure 
 versus temperature phase diagram for isochores between $0.05$ and $0.50$ in solid gray lines. In 
(b) we show the lateral diffusion coefficient as function of time for fixed temperatures 
of $T = 0.150$, $0.200$, $0.250$, $0.300$ and $0.350$.}
 \label{phase-diagram}
 \end{figure}

With this informations, we can draw the $P_{\parallel} - T$ phase diagram for this system,
shown in figure~\ref{phase-diagram}(a). 
We should address that the phase diagram is qualitative,
based on direct observation of the various assembled structures, the $v_{acf}$, the $g_{||}(r_{||})$ and lateral mean square displacement.
The regions where the distinct self-assembled structures were observed are indicated in
the phase-diagram. In addition to the anomalous diffusion, the system also shows diffusion anomaly.
This anomaly is characterized by the maximum and minimum in the curve of the lateral diffusion coefficient $D_{||}$ 
as function of density at constant temperature. The figure~\ref{phase-diagram}(b) shows these curves, with
the diffusion extrema indicated. This extrema is also shown in the phase diagram. 
Notice that the anomaly region occurs in the fluid and in the SM phase, indicating that 
the anomaly can be observed in spherical micelles. In this way, it is possible to 
construct spherical self-assembled structure that will diffuse faster when compressed.
As usual, the diffusion anomaly is explained using the competition between the two scales.
The graphic showing this was omitted for simplicity, since this result is well known
and discussed in the literature~\cite{Ol06a, BoK15c, Bordin16a}.
 
Another interesting difference between the Brownian and the molecular system 
$P_{\parallel} - T$ phase diagram is the absence of a melting induced by the
density increase in the SEM phase~\cite{BoK16a}. This suggests that the melting
scenario for this Janus nanoparticles is strongly affected by the solvent. As well,
the fluid phase structure is also affected. This is evident once for the 
molecular system we have not observed the density anomaly, only the diffusion
anomaly. For the Brownian system, the density anomalous region ended in the 
melting induced by density increase region. Therefore, this comparison
between the Browinian and molecular system has shown that not only the dynamical behavior
are distinct, what was expected due the thermostats characteristics, 
but also the structural, assembly and thermodynamic
properties of these nanoparticles are strongly affected.

\section{Conclusion}
\label{Conclu}

We reported the study of Janus nanoparticles confined in a thin film. Here, the effects of Brownian motion were 
removed, and we studied the molecular system without solvent. 
This system has special interest in the design of
new material using the confinement to control the self-assembled structures. We have found a rich variety
of aggregates and micelles, including structures not observed in the bulk system or in the Brownian system. 
More than that, our results
show that the more structured micelle has an anomalous diffusion, with a superdiffusive regime
related to a maximum in the translational order parameter. We have shown that this anomalous diffusion
is related to the competition between the two length scales. As well, the system have diffusion anomaly,
where the diffusion constant increases with the density increases. 
These results show that materials which can be modeled by two length scale potentials have an interesting
and peculiar behavior. New studies on this system are in progress, as anisotropy effects.

\section{Acknowledgments}

We thank the Brazilian agency CNPq for the financial support.





\providecommand*{\mcitethebibliography}{\thebibliography}
\csname @ifundefined\endcsname{endmcitethebibliography}
{\let\endmcitethebibliography\endthebibliography}{}

\end{document}